# On Purely Macroscopic Theory of Rigid Semiconductors Constructed from Multi-Continuum Model

Jiashi Yang (jyang1@unl.edu)
Department of Mechanical and Materials Engineering
University of Nebraska-Lincoln, Lincoln, NE 68588-0526, USA

**Abstract**
A complete and systematic derivation of the purely macroscopic theory of rigid semiconductors is given. It is based on a five-continuum model of charged and interpenetrating continua, and the applications of the relevant laws of physics to the continua. The theory is within the quasistatic approximation of electrostatics and is limited to nonmagnetizable materials.

## 1. Introduction

Common treatment of semiconductors employs both the macroscopic theory of electrostatics and the microscopic quantum-statistical mechanics for energy bands as well as electron and hole distributions [1,2]. This may be viewed as a mixed approach with both macroscopic and microscopic components, and is different from the purely macroscopic approach of continuum mechanics or continuum physics. In fact, purely macroscopic theories of elastic [3,4] and rigid [5–7] semiconductors exist, but somehow have received limited attention. For example, Ref. [5] which is the major reference for this short article has been cited 27 times only since its publication in 1980. One reason may be that a big portion of [5] was devoted to analyzing an interface problem while the presentation of the theory itself was somewhat brief, omitting certain details such as the derivations of electric polarization and dielectric loss, etc., and that the basic derivations of equations appear in separate places in [5] either as part of the text or in an appendix. The purely macroscopic theory in [5] has certain advantages. For example, the electron and hole fluids in the theory have partial pressures which are useful in prescribing boundary conditions. A complete and systematic presentation of the macroscopic theory of rigid semiconductors in [5] is given below. The main purpose is to promote the awareness of [5] and help the understanding of it as well as to facilitate the transition from [5] on rigid bodies to the more involved macroscopic theory of elastic semiconductors in [3]. The macroscopic and multi-continuum models used in [3–7] for semiconductors are in fact very general and have been used to construct various other macroscopic theories of electromagnetoelastic solids in [8–17].

## 2. Five-Continuum Mode

For constructing a macroscopic theory of rigid and nonmagnetizable semiconductors, five continua are needed [5]. They are the lattice, bound charge, impurity, electron and hole continua. The position of a typical material point of the lattice continuum is denoted by **y**. The lattice is rigid and fixed. Its mass density $\rho$ is a constant. Its charge density is $\mu^l$. The impurity continuum with a charge density of $\mu^i$ is fixed to the lattice. The electrons and holes are represented by two massless fluids flowing through the lattice with charge densities of $\mu^e$ and $\mu^h$, respectively. The bound charge continuum is massless and its charge density is $\mu^b$. The bound charge continuum is permitted to displace with respect to the lattice continuum by an infinitesimal displacement field $\mathbf{\eta}(\mathbf{y},t)$. The sum of the lattice and bound charges at the following corresponding positions gives the lattice residual charge as

$$\mu^l(\mathbf{y}) + \mu^b(\mathbf{y} + \mathbf{\eta}) = \mu^r(\mathbf{y}). \tag{1}$$

It is assumed that $\mathbf{\eta}(\mathbf{y},t)$ preserves the volume of the bound charge continuum [8], i.e.,

$$\eta_{k,k} = 0. \tag{2}$$

The polarization and electric displacement vectors are defined by [8]

$$\begin{aligned}\mathbf{P} &= \mu^l(\mathbf{y})(-\mathbf{\eta}) = \mu^b(\mathbf{y}+\mathbf{\eta})\mathbf{\eta} \cong \mu^b(\mathbf{y})\mathbf{\eta}, \\ \mathbf{D} &= \varepsilon_0 \mathbf{E} + \mathbf{P},\end{aligned} \tag{3}$$

where only linear terms of $\mathbf{\eta}$ are kept. The electron, hole and impurity continua have charge sources which are assumed to satisfy [5]

$$\gamma^e + \gamma^h + \gamma^i = 0. \tag{4}$$

We denote the total charge and current densities by

$$\mu = \mu^r + \mu^e + \mu^h + \mu^i, \tag{5}$$

$$\mathbf{J} = \mu^e \mathbf{v}^e + \mu^h \mathbf{v}^h = \mathbf{J}^e + \mathbf{J}^h, \tag{6}$$

where $\mathbf{J}^e$ and $\mathbf{J}^h$ are electron and hole current densities.

### 3. Integral Balance Laws

The relevant integral balance laws are the conservation of various charges (continuity equations):

$$\frac{\partial}{\partial t}\int_v \mu^i dv = \int_v \gamma^i dv, \tag{7}$$

$$\frac{\partial}{\partial t}\int_v \mu^e dv = -\int_s \mathbf{n}\cdot\mathbf{v}^e \mu^e ds + \int_v \gamma^e dv, \tag{8}$$

$$\frac{\partial}{\partial t}\int_v \mu^h dv = -\int_s \mathbf{n}\cdot\mathbf{v}^h \mu^h ds + \int_v \gamma^h dv, \tag{9}$$

where $\mathbf{v}^e$ and $\mathbf{v}^h$ are the velocity fields of the electron and hole fluids. The addition of Eqs. (7)–(9) yields

$$\frac{\partial}{\partial t}\int_v \mu dv = -\int_s \mathbf{n}\cdot\mathbf{J} ds. \tag{10}$$

Obviously, Eq. (10) is not independent to Eqs. (7)–(9). For the quasistatic electric field under consideration, we have

$$\int_s \mathbf{n}\cdot\varepsilon_0 \mathbf{E} ds = \int_v (\mu^l + \mu^b + \mu^e + \mu^h + \mu^i) dv, \tag{11}$$

$$\int_l \mathbf{E}\cdot d\mathbf{l} = 0. \tag{12}$$

The electron and hole fluids are massless and hence are without momenta. Their pressure fields are denoted by $p^e$ and $p^h$. They interact with the lattice through effective local electric fields $\mathbf{E}^e$ and $\mathbf{E}^h$, respectively. The linear momentum equations for the fluids are

$$\int_s -p^e \mathbf{n} ds + \int_v \mu^e \left(\mathbf{E} + \mathbf{E}^e\right) dv = 0, \tag{13}$$

$$\int_s -p^h \mathbf{n} ds + \int_v \mu^h (\mathbf{E} + \mathbf{E}^h) dv = 0. \tag{14}$$

The energy equation and entropy inequality for the combined continuum of all five constituents together in a fixed region $v$ may be written as

$$\frac{\partial}{\partial t} \int_v (\rho \varepsilon + \mu^e \varepsilon^e + \mu^h \varepsilon^h) dv = \int_s \left( -p^e \mathbf{n} \cdot \mathbf{v}^e - p^h \mathbf{n} \cdot \mathbf{v}^h \right.$$
$$\left. - \mathbf{n} \cdot \mu^e \varepsilon^e \mathbf{v}^e - \mathbf{n} \cdot \mu^h \varepsilon^h \mathbf{v}^h - \mathbf{n} \cdot \mathbf{q} \right) ds \tag{15}$$
$$+ \int_v \left( \mu^e \mathbf{E} \cdot \mathbf{v}^e + \mu^h \mathbf{E} \cdot \mathbf{v}^h + \mathbf{E} \cdot \frac{\partial \mathbf{P}}{\partial t} + \rho r \right) dv,$$

$$\frac{\partial}{\partial t} \int_v \rho \eta dv \geq \int_v \frac{\rho r}{\theta} dv - \int_s \frac{\mathbf{q} \cdot \mathbf{n}}{\theta} ds, \tag{16}$$

where $\varepsilon$, $\varepsilon^e$ and $\varepsilon^h$ are the internal energy densities of the lattice, electron fluid and hole fluid, respectively. $\eta$ is the entropy density of the combined continuum. $\mathbf{q}$ is the heat flux vector and $r$ the body heat source. The work done by the electric field $\mathbf{E}$ on the polarization $\mathbf{P}$ is the dot product of $\mathbf{E}$ and $\partial \mathbf{P}/\partial t$ [8].

## 4. Differential Balance Laws

With the use of the divergence theorem and Stokes' theorem, the differential forms of Eqs. (7)–(16) can be obtained in a standard way as

$$\frac{\partial \mu^i}{\partial t} = \gamma^i, \tag{17}$$

$$\frac{\partial \mu^e}{\partial t} + \nabla \cdot (\mu^e \mathbf{v}^e) = \gamma^e, \tag{18}$$

$$\frac{\partial \mu^h}{\partial t} + \nabla \cdot (\mu^h \mathbf{v}^h) = \gamma^h, \tag{19}$$

$$\frac{\partial \mu}{\partial t} + \nabla \cdot \mathbf{J} = 0, \tag{20}$$

$$\nabla \cdot \mathbf{D} = \mu, \tag{21}$$

$$\nabla \times \mathbf{E} = 0, \quad \mathbf{E} = -\nabla \varphi, \tag{22}$$

$$-\nabla p^e - \mu^e \nabla \varphi + \mu^e \mathbf{E}^e = 0, \tag{23}$$

$$-\nabla p^h - \mu^h \nabla \varphi + \mu^h \mathbf{E}^h = 0, \tag{24}$$

$$\rho \frac{\partial \varepsilon}{\partial t} + \frac{\partial}{\partial t} (\mu^e \varepsilon^e + \mu^h \varepsilon^h) + \nabla \cdot (p^e \mathbf{v}^e + p^h \mathbf{v}^h + \mu^e \varepsilon^e \mathbf{v}^e + \mu^h \varepsilon^h \mathbf{v}^h)$$
$$= \mu^e \mathbf{E} \cdot \mathbf{v}^e + \mu^h \mathbf{E} \cdot \mathbf{v}^h + \mathbf{E} \cdot \frac{\partial \mathbf{P}}{\partial t}, \tag{25}$$

$$\rho \frac{\partial \eta}{\partial t} \geq \frac{\rho r}{\theta} - \left( \frac{q_i}{\theta} \right)_{,i}, \tag{26}$$

where $\varphi$ is the usual electrostatic potential. When the integral balance laws are applied to an interface or a boundary surface of a semiconductor body, they lead to jump or boundary conditions. At interfaces or boundaries of semiconductors there may be surface charges or currents which contribute to the jump or boundary conditions there [3].

## 5. Constitutive Relations

With the use of the continuity equations of various charges and the linear momentum equations for the electron and hole fluids, the energy equation in Eq. (25) can be written as

$$\rho \frac{\partial \varepsilon}{\partial t} + \mu^e \frac{d^e \varepsilon^e}{dt} + \mu^h \frac{d^h \varepsilon^h}{dt} - \frac{p^e}{\mu^e} \frac{d^e \mu^e}{dt} - \frac{p^h}{\mu^h} \frac{d^h \mu^h}{dt}$$
$$= -\mu^e \mathbf{E}^e \cdot \mathbf{v}^e - \mu^h \mathbf{E}^h \cdot \mathbf{v}^h + \mathbf{E} \cdot \frac{\partial \mathbf{P}}{\partial t} \qquad (27)$$
$$- \gamma^e \left( \frac{p^e}{\mu^e} + \varepsilon^e \right) - \gamma^h \left( \frac{p^h}{\mu^h} + \varepsilon^h \right) + \rho r - q_{i,i},$$

where two material derivatives, $d^e/dt$ and $d^h/dt$, following the electron fluid and the hole fluid have been introduced. Eliminating $r$ from Eqs. (26) and (27), we obtain the Clausius–Duhem inequality as

$$\rho \theta \frac{\partial \eta}{\partial t} - \rho \frac{\partial \varepsilon}{\partial t} - \mu^e \frac{d^e \varepsilon^e}{dt} - \mu^h \frac{d^h \varepsilon^h}{dt}$$
$$+ \frac{p^e}{\mu^e} \frac{d^e \mu^e}{dt} + \frac{p^h}{\mu^h} \frac{d^h \mu^h}{dt} - \mu^e \mathbf{E}^e \cdot \mathbf{v}^e - \mu^h \mathbf{E}^h \cdot \mathbf{v}^h + \mathbf{E} \cdot \frac{\partial \mathbf{P}}{\partial t} \qquad (28)$$
$$- \gamma^e \left( \frac{p^e}{\mu^e} + \varepsilon^e \right) - \gamma^h \left( \frac{p^h}{\mu^h} + \varepsilon^h \right) - \frac{q_i \theta_{,i}}{\theta} \geq 0.$$

A free energy $F$ can be introduced through

$$\rho F = \rho \varepsilon - \mathbf{E} \cdot \mathbf{P} - \rho \eta \theta. \qquad (29)$$

Then the energy equation and Clausius–Duhem inequality in Eq. (27) and (28) take the following form:

$$\rho \frac{\partial F}{\partial t} + \rho \theta \frac{\partial \eta}{\partial t} + \rho \eta \frac{\partial \theta}{\partial t} + \mu^e \frac{d^e \varepsilon^e}{dt} + \mu^h \frac{d^h \varepsilon^h}{dt} - \frac{p^e}{\mu^e} \frac{d^e \mu^e}{dt} - \frac{p^h}{\mu^h} \frac{d^h \mu^h}{dt}$$
$$= -\mu^e \mathbf{E}^e \cdot \mathbf{v}^e - \mu^h \mathbf{E}^h \cdot \mathbf{v}^h - \mathbf{P} \cdot \frac{\partial \mathbf{E}}{\partial t} \qquad (30)$$
$$- \gamma^e \left( \frac{p^e}{\mu^e} + \varepsilon^e \right) - \gamma^h \left( \frac{p^h}{\mu^h} + \varepsilon^h \right) + \rho r - q_{i,i},$$

$$-\rho \frac{\partial F}{\partial t} - \rho \eta \frac{\partial \theta}{\partial t} - \mu^e \frac{d^e \varepsilon^e}{dt} - \mu^h \frac{d^h \varepsilon^h}{dt} + \frac{p^e}{\mu^e} \frac{d^e \mu^e}{dt} + \frac{p^h}{\mu^h} \frac{d^h \mu^h}{dt}$$
$$- \mu^e \mathbf{E}^e \cdot \mathbf{v}^e - \mu^h \mathbf{E}^h \cdot \mathbf{v}^h - \mathbf{P} \cdot \frac{\partial \mathbf{E}}{\partial t} \qquad (31)$$
$$- \gamma^e \left( \frac{p^e}{\mu^e} + \varepsilon^e \right) - \gamma^h \left( \frac{p^h}{\mu^h} + \varepsilon^h \right) - \frac{q_i \theta_{,i}}{\theta} \geq 0.$$

Consider the case when

$$F = F(\mathbf{E}; \theta), \quad \mathbf{P} = \mathbf{P}^R + \mathbf{P}^D,$$
$$\varepsilon^e = \varepsilon^e(\mu^e), \quad \varepsilon^h = \varepsilon^h(\mu^h), \tag{32}$$

where the possible dependence of the internal energy densities of electrons and holes on $\theta$ is not considered. $\mathbf{P}^R$ and $\mathbf{P}^D$ are for the reversible and dissipative parts of the electric polarization. The reversible constitutive relations satisfy

$$-\rho\frac{\partial F}{\partial t} - \rho\eta\frac{\partial \theta}{\partial t} - \mu^e\frac{d^e \varepsilon^e}{dt} - \mu^h\frac{d^h \varepsilon^h}{dt}$$
$$+ \frac{p^e}{\mu^e}\frac{d^e \mu^e}{dt} + \frac{p^h}{\mu^h}\frac{d^h \mu^h}{dt} - \mathbf{P}^R \cdot \frac{\partial \mathbf{E}}{\partial t} = 0, \tag{33}$$

or

$$\left(P_k^R + \rho\frac{\partial F}{\partial E_k}\right)\frac{\partial E_k}{\partial t} + \rho\left(\eta + \frac{\partial F}{\partial \theta}\right)\frac{\partial \theta}{\partial t}$$
$$+ \left(\mu^e \frac{\partial \varepsilon^e}{\partial \mu^e} - \frac{p^e}{\mu^e}\right)\frac{d^e \mu^e}{dt} + \left(\mu^h \frac{\partial \varepsilon^h}{\partial \mu^h} - \frac{p^h}{\mu^h}\right)\frac{d^h \mu^h}{dt} = 0. \tag{34}$$

Equation (34) implies the following reversible constitutive relations:

$$P_k^R = -\rho\frac{\partial F}{\partial E_k}, \quad \eta = -\frac{\partial F}{\partial \theta},$$
$$p^e = \left(\mu^e\right)^2 \frac{\partial \varepsilon^e}{\partial \mu^e}, \quad p^h = \left(\mu^h\right)^2 \frac{\partial \varepsilon^h}{\partial \mu^h}. \tag{35}$$

What are left from the energy equation and Clausius–Duhem inequality in Eqs. (30) and (31) are

$$\rho\theta\frac{\partial \eta}{\partial t} = -\mu^e \mathbf{E}^e \cdot \mathbf{v}^e - \mu^h \mathbf{E}^h \cdot \mathbf{v}^h - \mathbf{P}^D \cdot \frac{\partial \mathbf{E}}{\partial t}$$
$$-\gamma^e\left(\frac{p^e}{\mu^e} + \varepsilon^e\right) - \gamma^h\left(\frac{p^h}{\mu^h} + \varepsilon^h\right) + \rho r - q_{i,i}, \tag{36}$$

$$-\mu^e \mathbf{E}^e \cdot \mathbf{v}^e - \mu^h \mathbf{E}^h \cdot \mathbf{v}^h - \mathbf{P}^D \cdot \frac{\partial \mathbf{E}}{\partial t}$$
$$-\gamma^e\left(\frac{p^e}{\mu^e} + \varepsilon^e\right) - \gamma^h\left(\frac{p^h}{\mu^h} + \varepsilon^h\right) - \frac{q_i \theta_{,i}}{\theta} \geq 0. \tag{37}$$

Equation (36) is the heat or dissipation equation. The dissipative constitutive relations may assume, for example, the following form [5]:

$$\mathbf{E}^e = \mathbf{E}^e(\mu^e, \mathbf{v}^e, \mathbf{E}, \theta), \quad \mathbf{E}^h = \mathbf{E}^h(\mu^h, \mathbf{v}^h, \mathbf{E}, \theta),$$
$$\mathbf{P}^D = \mathbf{P}^D(\mu^e, \mu^h, \mathbf{v}^e, \mathbf{v}^h, \mathbf{E}, \theta),$$
$$\gamma^e = \gamma^e(\mu^e, \mu^h, \mathbf{v}^e, \mathbf{v}^h, \mathbf{E}, \theta), \tag{38}$$
$$\gamma^h = \gamma^h(\mu^e, \mu^h, \mathbf{v}^e, \mathbf{v}^h, \mathbf{E}, \theta),$$
$$\mathbf{q} = \mathbf{q}(\mu^e, \mu^h, \mathbf{v}^e, \mathbf{v}^h, \mathbf{E}, \theta, \theta_{,k}),$$

which are restricted by Eq. (37).

## 6. Chemical Potential

It is convenient to introduce the following chemical potentials:
$$\varphi^e = \frac{\partial(\mu^e \varepsilon^e)}{\partial \mu^e}, \quad \varphi^h = \frac{\partial(\mu^h \varepsilon^h)}{\partial \mu^h}, \tag{39}$$

which have the following functional forms:
$$\varphi^e = \varphi^e(\mu^e), \quad \varphi^h = \varphi^h(\mu^h). \tag{40}$$

From Eq. (39) and the last two equations in Eq. (35), we obtain
$$\varphi^e = \frac{\partial(\mu^e \varepsilon^e)}{\partial \mu^e} = \varepsilon^e + \mu^e \frac{\partial \varepsilon^e}{\partial \mu^e} = \varepsilon^e + \frac{1}{\mu^e}(\mu^e)^2 \frac{\partial \varepsilon^e}{\partial \mu^e} = \varepsilon^e + \frac{p^e}{\mu^e}, \tag{41}$$

$$\frac{p^h}{\mu^h} + \varepsilon^h = \varphi^h. \tag{42}$$

In terms of the chemical potentials, the dissipation equation and the Clausius–Duhem inequality in Eqs. (36) and (37) become
$$\rho\theta \frac{\partial \eta}{\partial t} = -\mu^e \mathbf{E}^e \cdot \mathbf{v}^e - \mu^h \mathbf{E}^h \cdot \mathbf{v}^h - \mathbf{P}^D \cdot \frac{\partial \mathbf{E}}{\partial t} - \gamma^e \varphi^e - \gamma^h \varphi^h + \rho r - q_{i,i}, \tag{43}$$

$$-\mu^e \mathbf{E}^e \cdot \mathbf{v}^e - \mu^h \mathbf{E}^h \cdot \mathbf{v}^h - \mathbf{P}^D \cdot \frac{\partial \mathbf{E}}{\partial t} - \gamma^e \varphi^e - \gamma^h \varphi^h - \frac{q_i \theta_{,i}}{\theta} \geq 0. \tag{44}$$

From Eq. (41), we have
$$\varphi_{,i}^e = \frac{\partial \varepsilon^e}{\partial \mu^e} \mu_{,i}^e + \frac{p_{,i}^e \mu^e - p^e \mu_{,i}^e}{(\mu^e)^2} = \frac{\partial \varepsilon^e}{\partial \mu^e} \mu_{,i}^e + \frac{p_{,i}^e}{\mu^e} - \frac{p^e}{(\mu^e)^2} \mu_{,i}^e$$
$$= \left(\frac{\partial \varepsilon^e}{\partial \mu^e} - \frac{p^e}{(\mu^e)^2}\right) \mu_{,i}^e + \frac{p_{,i}^e}{\mu^e} = \frac{p_{,i}^e}{\mu^e}. \tag{45}$$

Hence
$$\frac{1}{\mu^e} \nabla p^e = \nabla \varphi^e, \quad \frac{1}{\mu^h} \nabla p^h = \nabla \varphi^h. \tag{46}$$

Then the linear momentum equations for the electron and hole fluids in Eqs. (23) and (24) can be written as
$$-\nabla(\varphi^e + \varphi) + \mathbf{E}^e = 0, \tag{47}$$
$$-\nabla(\varphi^h + \varphi) + \mathbf{E}^h = 0. \tag{48}$$

## 7. Example of Constitutive Relations

As an example of constitutive relations for electron and hole currents, consider the common constitutive relations below [1,2]:

$$\mathbf{E}^e = \mathbf{v}^e / m^e, \quad \mathbf{E}^h = -\mathbf{v}^h / m^h, \tag{49}$$

where $m^e$ and $m^h$ are the mobility of electrons and holes. Then

$$\mathbf{J}^e = \mu^e \mathbf{v}^e = m^e \mu^e \mathbf{E}^e = m^e \mu^e (-\mathbf{E} + \nabla \varphi^e) = m^e \mu^e \left( -\mathbf{E} + \frac{\partial \varphi^e}{\partial \mu^e} \nabla \mu^e \right)$$
$$= -m^e \mu^e \mathbf{E} + m^e \mu^e \frac{\partial \varphi^e}{\partial \mu^e} \nabla \mu^e = -m^e \mu^e \mathbf{E} - D^e \nabla \mu^e, \tag{50}$$

where Eq. (47) has been used. Hence

$$\mathbf{J}^e = -m^e \mu^e \mathbf{E} - D^e \nabla \mu^e, \tag{51}$$

where the diffusion constant is defined by

$$D^e = -m^e \mu^e \frac{\partial \varphi^e}{\partial \mu^e}. \tag{52}$$

Similarly,

$$\mathbf{J}^h = m^h \mu^h \mathbf{E} - D^h \nabla \mu^h, \tag{53}$$

$$D^h = m^h \mu^h \frac{\partial \varphi^h}{\partial \mu^h}. \tag{54}$$

## References

[1] R.F. Pierret, *Semiconductor Device Fundamentals*, Pearson, Uttar Pradesh, India, 1996.
[2] S.M. Sze, *Physics of Semiconductor Devices*, John Wiley & Sons, New York, 1981.
[3] H.G. de Lorenzi and H.F. Tiersten, On the interaction of the electromagnetic field with heat conducting deformable semiconductors, *J. Math. Phys.*, 16, 938–957, 1975.
[4] M.F. McCarthy and H.F. Tiersten, On integral forms of the balance laws for deformable semiconductors, *Archive for Rational Mechanics and Analysis*, 68, 27–36, 1978.
[5] M.G. Ancona and H.F. Tiersten, Fully macroscopic description of bounded semiconductors with an application to the Si-SiO$_2$ interface, *Phys. Rev. B*, 22, 6104–6119, 1980.
[6] M.G. Ancona and H.F. Tiersten, Fully macroscopic description of electrical conduction in metal-insulator-semiconductor structures, *Phys. Rev. B*, 27, 7018–7045, 1983.
[7] M.G. Ancona and H.F. Tiersten, Macroscopic physics of the silicon inversion layer, *Phys. Rev. B*, 35, 7959–7965, 1987.
[8] H.F. Tiersten, On the nonlinear equations of thermoelectroelasticity, *Int. J. Engng Sci.*, 9, 587–604, 1971.
[9] H.F. Tiersten, On the interaction of the electromagnetic field with deformable solid continua, in: *Electromagnetomechanical Interactions in Deformable Solids and Structures*, Y. Yamamoto and K. Miya, ed., North-Holland, 1987, pp. 277–284.

[10] H.F. Tiersten, An extension of the London equations of superconductivity, *Physica*, 37, 504-538, 1967.

[11] H.F. Tiersten, Coupled magnetomechanical equation for magnetically saturated insulators, *J. Math. Phys.*, 5, 1298–1318, 1964.

[12] H.F. Tiersten and C.F. Tsai, On the interaction of the electromagnetic field with heat conducting deformable insulators, *J. Math. Phys.*, 13, 361–378, 1972.

[13] J.S. Yang, *An Introduction to the Theory of Piezoelectricity*, 2nd ed., World Scientific, Singapore, 2018.

[14] J.S. Yang, An alternative derivation of the Landau–Lifshitz–Gilbert equation for saturated ferromagnets, *arXiv*, 2305.18232, 2023.

[15] J.S. Yang, A macroscopic theory of saturated ferromagnetic conductors, *arXiv*, 2306.11525, 2023.

[16] J.S. Yang, A continuum theory of elastic-ferromagnetic conductors, *arXiv*, 2307.16669, 2023.

[17] J.S. Yang, *Theory of Electromagnetoelasticity*, World Scientific, Singapore, 2024.